\documentclass[aps,prd,preprint,groupedaddress,nofootinbib]{revtex4-1}

\usepackage{amssymb}
\usepackage{hyperref}

\begin{document}

\title{  Gauge Theory Formulations for Continuous  \\ and Higher Spin Fields}

%% %simple case: 2 authors, same institution
\author{Victor O. Rivelles}
%% \author{and A. Nother Author}
\affiliation{ Instituto de F\'{i}sica, Universidade de S\~ao Paulo,\\C. Postal 66318, 05314-970 S\~ao Paulo, SP, Brazil}

% more complex case: 4 authors, 3 institutions, 2 footnotes
%\author[a,b,1]{F. Irst,\note{Corresponding author.}}
%\author[c]{S. Econd,}
%\author[a,2]{T. Hird\note{Also at Some University.}}
%\author[a,2]{and Fourth}

% The "\note" macro will give a warning: "Ignoring empty anchor..."
% you can safely ignore it. 

%\affiliation[a]{One University,\\some-street, Country}
%\affiliation[b]{Another University,\\different-address, Country}
%\affiliation[c]{A School for Advanced Studies,\\some-location, Country}

% e-mail addresses: one for each author, in the same order as the authors
\email{rivelles@fma.if.usp.br}
%\emailAdd{second@asas.edu}
%\emailAdd{third@one.univ}
%\emailAdd{fourth@one.univ}

\begin{abstract}We consider a gauge theory action for continuous spin particles formulated in a spacetime enlarged by an extra coordinate recently proposed  by Schuster and Toro. It requires one scalar gauge field and has two local symmetries. We show that the local symmetries are reducible in the sense that the parameters also have a local symmetry. Using reducibility we get an action which has two scalar gauge fields and a reducible but simpler local symmetry. We then show how this action and equations of motion are related to previously proposed formulations for continuous spin particles and higher spin theories. We also discuss the physical contents of each formulation to reveal the physical degrees of freedom.  
\end{abstract}


\maketitle

\section{Introduction}

The less known particle type allowed by special relativity and quantum mechanics is the continuous spin particle (CSP). Along with massive and massless particles of integer and half-integer spins they constitute the irreducible representations of the Poincar\'e group \cite{Wigner:1939cj}. CSPs are massless states labelled by a real parameter $\rho$, its continuous spin, and comprises infinitely many helicity states which mix with each other under Lorentz transformations. There is a bosonic representation where all helicities are integer and a fermionic one with all  helicities being half-integer. When $\rho=0$ the helicity states decouple of each other and reduce to a set of ordinary massless states in which each helicity appears once giving rise to a higher spin (HS) theory having all integer (or half-integer) helicities present. CSPs are largely ignored not only because they are not found in Nature but also because even its free quantum formulation is beset with problems \cite{Yngvason:1970fy,Iverson:1971hq,Chakrabarti:1971rz,Abbott:1976bb,Hirata:1977ss}. However, it was found recently that CSPs have covariant soft emission amplitudes which approach the amplitudes for ordinary low helicity particles ($0, \pm 1$ and $\pm 2$) at energies large compared with $\rho$ or in the non-relativistic regime \cite{Schuster:2013pxj,Schuster:2013vpr}. This led naturally to a search for an action principle, at least for the free case, and soon an action was proposed for a bosonic CSP\footnote{A previously proposed action \cite{Schuster:2013pta} described not a single CSP but a continuum of CSPs, with every value of $\rho$, making the coupling to a conserved current in the $\rho\rightarrow 0$ limit problematic.}  \cite{Schuster:2014hca}. As for other theories of massless particles it is a gauge theory. It can be coupled to currents which are consistent with no-go theorems for lower spins and with the covariant soft factors of \cite{Schuster:2013pxj,Schuster:2013vpr}. The equations of motion describe degrees of freedom with the expected polarization content of a single CSP. When $\rho$ vanishes the equations of motion reduce to the well known higher spin Fronsdal equations \cite{Fronsdal:1978rb} for all helicities. 

The action is formulated in an enlarged spacetime with the usual spacetime coordinates $x^\mu$ and an additional 4-vector coordinate $\eta^\mu$. The gauge field $\Psi(x,\eta)$ is a scalar field and it is assumed to be analytic in $\eta^\mu$. The action is given by \cite{Schuster:2014hca}
\begin{equation}\label{1}
	S = \frac{1}{2} \int d^4 x \,\, d^4\eta \left[ \delta^\prime(\eta^2+1) (\partial_x \Psi(\eta,x))^2 + \frac{1}{2} \delta(\eta^2+1) \left( (\partial_\eta \cdot \partial_x + \rho) \Psi(\eta,x) \right)^2 \right],
\end{equation}
where $\delta^\prime$ is the derivative of the delta function with respect to its argument. The spacetime metric is mostly minus. There are no terms with two derivatives of $\eta^0$. The action is invariant under Lorentz transformations and translations in $x^\mu$ but not translations in $\eta^\mu$. It is also invariant under two local transformations
\begin{equation}\label{2}
	\delta \Psi(\eta,x) = [ \eta\cdot\partial_x - \frac{1}{2} (\eta^2+1) (\partial_\eta \cdot \partial_x + \rho) ] \epsilon(\eta,x)  + \frac{1}{4} (\eta^2+1)^2 \chi(\eta,x),
\end{equation}
where $\epsilon(\eta,x)$ and $\chi(\eta,x)$ are the local parameters. As remarked in \cite{Schuster:2014hca} this represents a huge gauge freedom which was used to show that the action (\ref{1}) propagates only one CSP degree of freedom. Furthermore, when $\rho$ vanishes it was shown that the action describes HS fields for all integer helicities. 

We wish to remark that the local transformations (\ref{2}) are in fact reducible since
\begin{eqnarray}\label{3}
	\delta \epsilon &=& \frac{1}{2} (\eta^2+1) \Lambda(\eta,x), \\\label{3a}
	\delta \chi &=& (\partial_\eta\cdot\partial_x + \rho) \Lambda(\eta,x),
\end{eqnarray}
with $\Lambda(\eta,x)$ arbitrary leaves (\ref{2}) invariant. In Section \ref{s2} we will show that is possible to expand $\Psi(\eta,x)$ in powers of $\eta^2+1$ and keep only the first two terms. The action for these two fields also has a reducible but simpler local symmetry. This action is one of the main results of this paper and we will explore its consequences in the remaining sections. 

Earlier attempts to a formulation of CSPs involved the handling of the Bargmann-Wigner equations \cite{Bargmann:1948ck,Hirata:1977ss,Bengtsson:2013vra}, the proposal of covariant equations \cite{Iverson:1971hq,Abbott:1976bb} and also attempts to derive them from massive higher spin equations \cite{Bekaert:2005in}. No action was ever proposed before preventing the use of perturbation theory and coupling to ordinary fields to better understand the CSPs properties. Even so the CSP representations can be extended to higher dimensions and to the supersymmetric case \cite{Brink:2002zx} forming supermultiplets of the super-Poincar\'e group. Perturbative string theory does not allow CSPs providing one of the few model independent properties of low energy string theory \cite{Font:2013hia}. Tensionless strings, however, do propagate CSPs \cite{Savvidy:2003fx,Mourad:2006xk}. The proposal of the gauge invariant action (\ref{1}) for free  CSPs opens new doors to the understanding of this class of particles. 
%It can easily be extended to any dimension and can include fermions. 
It is  also relevant in $2+1$ dimensions providing a massless generalization of anyons \cite{Schuster:2014xja}.  
%If consistent interactions respecting the gauge symmetry can be found it could provide an interacting theory for HS massless states surpassing standard no-go theorems since it has a dimensionfull constant $\rho$. 
No self interactions or matter interaction are known presently but mass terms, for instance, are excluded  \cite{Schuster:2014hca}. Another important consequence of (\ref{1}) is that for $\rho=0$ it reduces to a sum of Fronsdal actions for massless higher spins for all helicities \cite{Schuster:2014hca} providing an alternative formulation for massless higher spin particles. 

The use of extra coordinates as a bookkeeping device has already been employed in some formulations of CSPs and HS fields. Starting with the Wigner conditions the authors of \cite{Bekaert:2005in} show that a limit of massive HS field equations results in the CSP equations of motion. They obtain a gauge invariant  equation for a single CSP in terms of a constrained field. To remove the constraint a compensator is introduced ending with a formulation with two gauge fields. No action giving these equations of motion was found. The field used in \cite{Bekaert:2005in} is composed of a totally symmetric tensor with all of its indices contracted with an auxiliary vector. It is natural to relate this extra vector with the coordinate $\eta$ of \cite{Schuster:2014hca}. We will show in Section \ref{s6} how to derive both formulations of \cite{Bekaert:2005in} starting with the action  derived in Section \ref{s2}. They also consider the case of a massless HS field and we will show how to derive their equations for $\rho=0$ in Section \ref{s3}. Our HS formulation corresponds to a previous proposal for HS fields which also considers two gauge fields making use of extra coordinates \cite{Segal:2001qq}. There is also a HS field formulation which uses an oscillator basis \cite{Metsaev:2011iz} instead of extra vectors. The indices of a totally symmetric HS field are contracted with creation operators forming a HS ket. An action can then be written in flat or AdS spaces. For the flat case the action reproduces Fronsdal equations. We will show in Section \ref{s4} how to relate the oscillator formalism action with the action derived in Section \ref{s2}. Finally in Section \ref{pc} we will discuss the physical contents of our formulation. We show that the Casimir operator has the correct value when acting on the fields and that the the fields carry all integer helicities just once as required for a CSP.

\section{Reducibility of the Gauge Transformations}\label{s2}

In order to explore the reducibility of the gauge transformations (\ref{2}) and to make contact with the results of \cite{Bekaert:2005in} we will choose a specific form for the $\eta$ dependence of $\Psi(\eta,x)$. We must first notice that the delta functions in (\ref{1}) are essentially restricting the $\eta$ dependence of $\Psi(\eta,x)$ to a hyperboloid in $\eta$-space so it is natural to assume the expansion 
\begin{equation}\label{5}
	\Psi(\eta,x) = \sum_{n=0}^{\infty} \frac{1}{n!} (\eta^2+1)^n \psi_n(\eta,x),
\end{equation}
where $\psi_n(\eta,x)$ are also scalar fields. Taking into account that the fields in \cite{Bekaert:2005in} depend on an extra vector contracted with HS fields we also assume that $\psi_n(\eta,x)$ are analytic in $\eta^\mu$
\begin{equation}\label{5a}
	\psi_n(\eta,x) = \sum_{s=0}^\infty \frac{1}{s!} \eta^{\mu_1}\dots\eta^{\mu_s} \psi^{(n,s)}_{\mu_1\cdots\mu_s}(x),
\end{equation}
where $\psi^{(n,s)}_{\mu_1\cdots\mu_s}(x)$ is a completely symmetric and unconstrained tensor field in spacetime. The choices (\ref{5}) and (\ref{5a}) are not unique. For instance, any trace of $\psi^{(n,s)}_{\mu_1\cdots\mu_s}(x)$ inserted in (\ref{5a}) will be proportional to $\eta^2$ and could be absorbed in $\psi_{n+1}(\eta,x)$. In general the local transformation 
\begin{equation}
	\delta \psi_n(\eta,x) = \sum_{p=1}^\infty \frac{n!}{(n+p)!} (\eta^2+1)^p \,\, \Xi_{n,n+p}(\eta,x) - \sum_{p=0}^{n-1} \Xi_{n,p}(\eta,x), \label{2.3}
\end{equation}
will leave $\Psi(\eta,x)$ invariant if $\Xi_{n,p}(\eta,x)$ is symmetric in $n$ and $p$.  This is not a gauge symmetry since it is not removing gauge degrees of freedom. It is just reshuffling them among $\psi_n(\eta,x)$ and it will be relevant for simplifying the $\eta$ structure of the equations of motion. All this does not seem to be a good idea since we are replacing the original field $\Psi$ by an infinite number of other fields $\psi_n$ but shortly we will see its relevance when we take into account the reducibility of the local transformations. 

Now let us consider in more detail the $\epsilon,\chi$ and $\Lambda$  transformations (\ref{2}-\ref{3a}). These parameters can also be expanded like (\ref{5}) and (\ref{5a}) so we find that (\ref{2}) reduces to 
\begin{equation}\label{8}
	\delta \psi_n = (1-n) \eta\cdot\partial_x \epsilon_n - \frac{1}{2} n \Delta \epsilon_{n-1} + \frac{1}{4} n(n-1) \chi_{n-2},
\end{equation}
where $\Delta = \partial_\eta \cdot \partial_x + \rho$, while (\ref{3}) and (\ref{3a}) become
\begin{eqnarray}\label{9}
	\delta \epsilon_n &=& \frac{1}{2} n \Lambda_{n-1}, \\
	\delta \chi_n &=& 2 \eta\cdot\partial_x \Lambda_{n+1} + \Delta \Lambda_n.
\end{eqnarray}

We can now use the $\Lambda$ symmetry to choose the gauge $\Lambda_{n-1} = -(2/n) \epsilon_n$ for $n\not= 0$ so that all $\epsilon_n$  with $n>0$ vanish while $\epsilon_0$ remains free. We can now fix the $\chi$ symmetry by choosing $\chi_{n-2} = -\frac{4}{n(n-1)} \psi_n$ for $n \ge 2$ so that $\psi_n=0$ for $n\ge 2$. Then only $\psi_0, \psi_1$ and $\epsilon_0$ are non-vanishing. We then find that (\ref{8}) and (\ref{9}) become
\begin{eqnarray}\label{10}
	\delta \psi_0 &=& \eta\cdot\partial_x \epsilon_0, \\ \label{10a}
	\delta \psi_1 &=& - \frac{1}{2} \Delta \epsilon_0,\\
	\delta \epsilon_0 &=& 0.
\end{eqnarray}
Then reducibility allowed us to eliminate all terms of $\epsilon$ except $\epsilon_0$ while the $\chi$ symmetry eliminated all terms of $\Psi$ except $\psi_0$ and $\psi_1$ which have a much more simple gauge transformation with an unconstrained parameter $\epsilon_0$. The gauge transformation is now irreducible. 

When the expansion (\ref{5}) is used in the action (\ref{1}) we get
\begin{equation}\label{1a}
	S =\frac{1}{2} \int d^4 x \,\, d^4\eta \left[ \delta^\prime(\eta^2+1) (\partial_x \psi_0)^2 + \frac{1}{2} \delta(\eta^2+1) \left[ \left( \Delta \psi_0 + 2 \eta\cdot\partial_x \psi_1 \right)^2 - 4 \partial_x \psi_0 \cdot \partial_x \psi_1 \right] \right].
\end{equation}
Notice that all $\psi_n$ with $n\ge 2$ have dropped out of the action because of the delta functions. The $\Lambda$ and $\chi$ symmetries were not used. So even without fixing the $\Lambda$ and $\chi$ symmetries the action knows about the reducibility of the original local transformations and only $\psi_0$ and $\psi_1$ remain at the end. 

Since we have found that the relevant fields are $\psi_0$ and $\psi_1$ we will consider from now on the action (\ref{1a}) as our starting point. The first step is to find out its local symmetries. We can easily verify that (\ref{1a}) is invariant under the gauge transformations (\ref{10}) and (\ref{10a}). The presence of delta functions means that the analogue of the original $\chi$ transformation of (\ref{2}) becomes $\delta \psi_0(\eta,x)= (\eta^2+1)^2 \chi_0(\eta,x)$ and $\delta \psi_1(\eta,x) = (\eta^2+1)\chi_1(\eta,x)$ with $\chi_0$ independent of $\chi_1$. So, together with the transformation (\ref{2.3}) that leaves $\Psi$ invariant, the action (\ref{1a}) is invariant under
\begin{eqnarray}
	\delta\psi_0(\eta,x) &=& \eta\cdot\partial_x \epsilon_0 + (\eta^2+1)^2 \chi_0(\eta,x) + (\eta^2+1) \Xi(\eta,x), \label{2.11}\\
	\delta\psi_1(\eta,x) &=& -\frac{1}{2} \Delta \epsilon_0 + (\eta^2+1) \chi_1(\eta,x) - \Xi(\eta,x). \label{2.12}
\end{eqnarray}
Reducibility of the above transformations now manifests itself as
\begin{eqnarray}
	\delta\Xi(\eta,x) &=& (\eta^2+1)\theta(\eta,x), \label{2.13} \\
	\delta\chi_0(\eta,x)  &=& - \theta(\eta,x), \label{2.14} \\
	\delta\chi_1(\eta,x) &=& \theta(\eta,x), \label{2.15}
\end{eqnarray}
and it essentially means that one of the local symmetry parameters $\chi_0, \chi_1$ or $\Xi$ is redundant. 

The equation of motion obtained by varying $\psi_0$ is\footnote{We could also derive the equations of motion directly from (\ref{1}) before making the expansion (\ref{5}) and get 
%\begin{equation}\label{4}
$	\delta^\prime(\eta^2+1) \Box_x \Psi - \frac{1}{2} \Delta \left( \delta(\eta^2+1) \Delta \right) \Psi = 0.$
%\end{equation}
Using now the expansion (\ref{5}) we get only (\ref{6}).}
\begin{equation}\label{6}
	\delta^\prime(\eta^2+1) \left[ \Box_x \psi_0 - \eta \cdot \partial_x \Delta \psi_0 - 2 (\eta \cdot \partial_x)^2 \psi_1 \right] - 2  \delta(\eta^2+1) \left[\Box_x \psi_1 + \frac{1}{2} \eta\cdot\partial_x \Delta \psi_1 + \frac{1}{4} \Delta^2 \psi_0 \right] = 0, 
\end{equation}
while varying $\psi_1$ yields
\begin{equation}\label{6a}
	\delta(\eta^2+1) \left[ \Box_x \psi_0 - \eta \cdot \partial_x \Delta \psi_0 - 2 (\eta \cdot \partial_x)^2 \psi_1 \right] = 0. 
\end{equation}
They are not independent since multiplying (\ref{6}) by $\eta^2+1$ we get (\ref{6a}). The delta functions in these equations are essentially restricting the field equations to the hyperboloid $\eta^2+1=0$. Solving the field equations on the hyperboloid is very cumbersome since $\eta^\mu$ is constrained. Instead we can use the reducibility of the gauge transformations to extend the field equations to all of $\eta$-space and only at the very end we go back to the hyperboloid. Proceeding in this way will also allow us to compare our formulation with previous ones. 

To this end let us call the first square bracket of (\ref{6}) as $A(\eta,x)$ and the second one as $B(\eta,x)$ 
\begin{eqnarray}
	A(\eta,x) &=&  \Box_x \psi_0 - \eta \cdot \partial_x \Delta \psi_0 - 2 (\eta \cdot \partial_x)^2 \psi_1, \label{2.16} \\
	B(\eta,x) &=& \Box_x \psi_1 + \frac{1}{2} \eta\cdot\partial_x \Delta \psi_1 + \frac{1}{4} \Delta^2 \psi_0, \label{2.16a}
\end{eqnarray}
so that (\ref{6}) and (\ref{6a}) become
\begin{eqnarray}
	\delta^\prime(\eta^2+1) A(\eta,x) - 2 \delta(\eta^2+1) B(\eta,x)&=&0, \label{2.17}\\
	\delta(\eta^2+1) A(\eta,x) &=&0. \label{2.18}
\end{eqnarray}
Notice that $A$ and $B$ are not completely independent since
\begin{equation}
	\Delta A = - 4 \eta\cdot\partial_x B. \label{2.18a}
\end{equation}
Notice also that $A$ and $B$ are invariant under $\epsilon_0$ gauge transformations but not under $\chi_0, \chi_1$ and $\Xi$ transformations since 
\begin{eqnarray}
	\delta A(\eta,x) &=& (\eta^2+1) \left[ (\Box_x - \eta\cdot\partial_x \Delta)\Xi + (\eta^2+1)(\Box_x - \eta\cdot\partial_x \Delta )\chi_0 \right. \nonumber \\ 
	& &\left.- 2 (\eta\cdot\partial_x)^2 (2\chi_0+\chi_1) \right],  \label{2.19}\\
	\delta B(\eta,x) &=& -\frac{1}{2} (\Box_x - \eta\cdot\partial_x\Delta)\Xi +  (\eta\cdot\partial_x)^2(2\chi_0+\chi_1)
	+ (\eta^2+1) [ (\Box_x  + 2\eta\cdot\partial_x \Delta )\chi_0   \nonumber \\ 
	& &  + (\Box_x +\frac{1}{2} \eta\cdot\partial_x \Delta )\chi_1 +  \frac{1}{4} \Delta^2 \Xi + \frac{1}{4} (\eta^2+1) \Delta^2 \chi_0 ]. \label{2.20}
	\end{eqnarray}
As expected, the above transformations are invariant under the $\theta$ transformations (\ref{2.13}-\ref{2.15}). We are then allowed to use these invariances to simplify and solve the delta function constraints on (\ref{2.17}) and (\ref{2.18}). 

Since (\ref{2.18}) is stating that $A(\eta,x)$ vanishes on the hyperboloid $\eta^2+1=0$ we will use the local symmetries in (\ref{2.19}) and (\ref{2.20}) to choose $A(\eta,x)=0$ in all of $\eta$ space. Under a $\chi_0$ and $\chi_1$ transformation  $A(\eta,x)=0$ implies that
\begin{equation}
	 (\eta^2+1)(\Box_x - \eta\cdot\partial_x \Delta )\chi_0 - 2 (\eta\cdot\partial_x)^2 (2\chi_0+\chi_1) = 0, \label{2.20a}
\end{equation}
outside the hyperboloid. We can now use the $\theta$ symmetry to choose $\chi_0=0$. It also reduces (\ref{2.20a}) to $(\eta\cdot\partial_x)^2 \chi_1=0$. This is a very restrictive equation. Expanding $\chi_1(\eta,x)$ as in (\ref{5a}) and going to momentum space it reads  $k_{(\mu_1} k_{\mu_2} \tilde{\chi}_{1\mu_3 \dots \mu_n)}(k)=0$ where $\tilde{\chi}_{1\mu_1\dots\mu_n}(k)$ is the Fourier transformed ${\chi}_{1\mu_1\dots\mu_n}(x)$. This means that either the momentum is constrained to vanish or that $\tilde{\chi}_{1\mu_1\dots\mu_n}(k)=0$. The first solution is not acceptable since it is fixing the momentum (this will be done when analysing the $\epsilon_0$ gauge transformations in Section \ref{pc}) and therefore $\chi_1(\eta,x)=0$ so that the $\chi_1$ symmetry is completely fixed. The only remaining symmetries are those generated by the $\Xi$ and $\epsilon_0$ transformations. Also having $A(\eta,x)=0$ means by (\ref{2.18a}) that $\eta\cdot\partial_x B=0$ and using the same argument as above we find that $B(\eta,x)=0$ as well. 

We have then found that $A(\eta,x)=B(\eta,x)=0$ so that the equations of motion (\ref{6}) and (\ref{6a}) have become 
\begin{eqnarray} \label{30}
	&\Box_x \psi_0 - \eta \cdot \partial_x \Delta \psi_0 - 2 (\eta \cdot \partial_x)^2 \psi_1 = 0, \\ \label{31}
	&\Box_x \psi_1 + \frac{1}{2} \eta\cdot\partial_x \Delta \psi_1 + \frac{1}{4} \Delta^2 \psi_0 =0,
\end{eqnarray}
and they still are invariant under
\begin{eqnarray}
	\delta\psi_0 &=& \eta\cdot\partial_x\epsilon_0 + (\eta^2+1) \Xi,  \label{2.25}\\
	\delta\psi_1 &=& -\frac{1}{2}\Delta\epsilon_0 - \Xi,  \label{2.26}
\end{eqnarray}
In this way we have taken into account all the effects of the delta functions so we can now analyse the consequences of (\ref{30}) and (\ref{31}). 

\section{Continuous Spin}\label{s6}

To make contact with the results of \cite{Bekaert:2005in} we will first consider the local $\Xi$ symmetry. It allow us to choose the gauge 
\begin{equation}
	\psi_0 - (\eta^2+1) \psi_1=0. \label{31.a}
\end{equation}
We now perform a Fourier transformation in $x^\mu$ and $\eta^\mu$
\begin{equation}\label{32}
	\psi_0(\eta,x) = \int d^4\omega \, d^4p \, e^{i\eta\cdot\omega + i p\cdot x} \tilde{\psi}_0(\omega,p),
\end{equation}
and similarly for $\psi_1$ and for $\epsilon_0$. We then find that (\ref{30}) and (\ref{31}) reduce to
\begin{eqnarray}\label{33}
	& p^2 \tilde{\psi}_0 + \frac{1}{2}(p\cdot\omega - \rho) p\cdot\partial_\omega \tilde{\psi}_0 + (p\cdot\partial_\omega)^2 \tilde{\psi}_1 = 0, \\ \label{33a}
	& p^2 \tilde{\psi}_1 - (p\cdot\omega - \rho) p\cdot\partial_\omega \tilde{\psi}_1 - \frac{1}{2} (p\cdot\omega - \rho)^2 \tilde{\psi}_0 =0,
\end{eqnarray}
and for the gauge transformations (\ref{2.25}) and (\ref{2.26}) we find
\begin{eqnarray}\label{34}
	\delta \tilde{\psi}_0 &=& -  p\cdot\partial_\omega \tilde{\epsilon}_0, \\ \label{34a}
	\delta \tilde{\psi}_1 &=& \frac{1}{2} (p\cdot\omega - \rho)\tilde{\epsilon}_0, 
\end{eqnarray}
 while the gauge choice (\ref{31.a}) becomes
\begin{equation}\label{35}
	\tilde{\psi}_0 = - (\Box_\omega -1) \tilde{\psi}_1.
\end{equation}
Using (\ref{35}) in (\ref{33a}) we find
\begin{equation}\label{36}
p^2 \tilde{\psi}_1 - (p\cdot\omega - \rho) p\cdot\partial_\omega \tilde{\psi}_1 + \frac{1}{2} (p\cdot\omega - \rho)^2 (\Box_\omega - 1) \tilde{\psi}_1 = 0,
\end{equation}
which is precisely equation (5.2) found in \cite{Bekaert:2005in}.
% after taking a limit in the massive higher spin equations. 
They also find that the field is constrained and we find the constraint after using (\ref{35}) in the field equation (\ref{33}) 
\begin{equation}\label{37}
	(p\cdot\omega - \rho)^2 (\Box_\omega - 1)^2 \tilde{\psi}_1 =0.
\end{equation}
Since we do not want to impose any condition on the momenta then $(\Box_\omega - 1)^2 \tilde{\psi}_1=0$ gives the trace condition (5.3) of \cite{Bekaert:2005in}. Finally we find that the gauge transformation (\ref{34a}) coincides with (5.5) of \cite{Bekaert:2005in}, while the consistency of the choice (\ref{35}) with (\ref{34}) yields 
\begin{equation}\label{38}
(p\cdot\omega - \rho) (\Box_\omega - 1)\tilde{\epsilon}_0 = 0.
\end{equation}
Not constraining the momenta means that $(\Box_\omega - 1)\tilde{\epsilon}_0=0$, the same condition found in equation (5.6) of \cite{Bekaert:2005in}. In this way we have reproduced the constrained formulation found in \cite{Bekaert:2005in} starting from the action (\ref{1a}). Notice that the constraint on $\tilde{\psi}_1$, that is, $(\Box_\omega - 1)^2 \tilde{\psi}_1=0$,  is in fact one of our equations of motion in our formulation. 

The trace condition on $\tilde{\psi}_1$ was removed in \cite{Bekaert:2005in} using a compensator field $\chi$. This corresponds to our formulation with both fields  $\psi_0$ and $\psi_1$. The field $\chi$ of \cite{Bekaert:2005in} can be introduced by a combination of $\tilde{\psi}_0$ and $\tilde{\psi}_1$ as
\begin{equation}\label{39}
	(p\cdot\omega - \rho) \chi = \tilde{\psi}_0 + (\Box_\omega - 1) \tilde{\psi}_1.
\end{equation}
 We then get from (\ref{33}) and (\ref{33a}) that 
\begin{eqnarray}\label{40}
	&& p^2 \tilde{\psi}_1 - (p\cdot\omega - \rho)p\cdot\partial_\omega\tilde{\psi}_1 + \frac{1}{2}(p\cdot\omega - \rho)^2 (\Box_\omega - 1)\tilde{\psi}_1 - \frac{1}{2} (p\cdot\omega - \rho)^3 \chi = 0, \\
	&& (\Box_\omega - 1)^2 \tilde{\psi}_1 - (p\cdot\omega - \rho) (\Box_\omega -1) \chi - 4 p\cdot\partial_\omega \chi = 0,
\end{eqnarray}
while the gauge transformation of $\chi$ becomes
\begin{equation}\label{41}
	\delta \chi = (\Box_\omega - 1) \tilde{\epsilon}_0.
\end{equation}
These correspond to equations (5.13-5.15) of \cite{Bekaert:2005in}. By choosing $\chi=0$  we recover from (\ref{39}) the constraint (\ref{35}), the trace condition of $\tilde{\psi}_1$ and $\tilde{\epsilon}_0$ besides (\ref{36}). Then we have shown that the equations proposed for CSPs in \cite{Bekaert:2005in} can be obtained from the action (\ref{1a}).

\section{Higher Spin Fields}\label{s3}

It was shown in \cite{Schuster:2014hca} that taking $\rho=0$ in (\ref{1}) reduces the action to a sum of Fronsdal actions for all integer helicities. Just setting $\rho=0$ in the equation for $\tilde{\psi}_0$ and $\tilde{\psi}_1$ of the last section does not lead us in an obvious way to any known formulation for HS theories. So let us go back to (\ref{1a}) and set $\rho=0$. Then the equations of motion (\ref{6}) and (\ref{6a}) reduce to 
\begin{eqnarray}\label{42}
	&\delta^\prime(\eta^2+1) \left[ \Box_x \psi_0 - \eta \cdot \partial_x  \,\, \partial_\eta \cdot \partial_x \psi_0 - 2 (\eta \cdot \partial_x)^2 \psi_1 \right] \nonumber \\  - 2 & \delta(\eta^2+1) \left[\Box_x \psi_1 + \frac{1}{2} \eta\cdot\partial_x  \,\, \partial_\eta \cdot \partial_x \psi_1 + \frac{1}{4} ( \partial_\eta\cdot\partial_x)^2 \psi_0 \right] = 0, 
\end{eqnarray}
and
\begin{equation}\label{42a}
	\delta(\eta^2+1) \left[ \Box_x \psi_0 - \eta \cdot \partial_x \,\, \partial_\eta \cdot \partial_x \psi_0 - 2 (\eta \cdot \partial_x)^2 \psi_1 \right] = 0,
\end{equation}
respectively. Using the same reasoning which lead to (\ref{30}) and (\ref{31})  we find 
\begin{eqnarray} \label{43}
	&\Box_x \psi_0 - \eta \cdot \partial_x \,\, \partial_\eta \cdot \partial_x \psi_0 - 2 (\eta \cdot \partial_x)^2 \psi_1 = 0, \\ \label{43a}
	&\Box_x \psi_1 + \frac{1}{2} \eta\cdot\partial_x  \,\, \partial_\eta \cdot \partial_x \psi_1 + \frac{1}{4} ( \partial_\eta\cdot\partial_x  )^2 \psi_0 =0.
\end{eqnarray}
The gauge transformations (\ref{2.25}) and (\ref{2.26}) are now
\begin{eqnarray}\label{44}
	\delta \psi_0 &=& \eta\cdot\partial_x \epsilon_0 + (\eta^2+1) \Xi, \\ \label{44a}
	\delta \psi_1 &=& - \frac{1}{2}  \partial_\eta \cdot \partial_x \epsilon_0 - \Xi.
\end{eqnarray}

To get the results of \cite{Bekaert:2005in} for the HS case we notice that $\Xi$ symmetry allow us to choose the gauge
\begin{equation}
\psi_1 + \frac{1}{4} \Box_\eta \psi_0=0, \label{45}
\end{equation}
which implies that 
\begin{equation}
	(1 + \eta\cdot\partial_\eta)\Xi + \frac{1}{4} (\eta^2+1) \Box_\eta \Xi = 0. \label{44b}
\end{equation} 
Then the equations of motion (\ref{43}) and (\ref{43a}) become
\begin{eqnarray}\label{46}
	& \Box_x \psi_0 - \eta \cdot \partial_x   \,\, \partial_\eta \cdot \partial_x  \psi_0 + \frac{1}{2} (\eta \cdot \partial_x)^2 \Box_\eta \psi_0 =0,  \\ \label{46a}
	& \Box_x \Box_\eta \psi_0 + \frac{1}{2} \eta\cdot\partial_x  \,\, \partial_\eta \cdot \partial_x  \Box_\eta \psi_0  - ( \partial_\eta\cdot\partial_x )^2 \psi_0  = 0.
\end{eqnarray}

Applying $\Box_\eta$ in (\ref{46}) and comparing with (\ref{46a}) gives $(\eta\cdot\partial_x)^2 \Box^2_\eta \psi_0 = 0$ 
which means that $\Box_\eta^2 \psi_0=0$, a double traceless condition on $\psi_0$. If we now take a $\Xi$ 
transformation of the double traceless condition we get $\Box_\eta^2 \delta\psi_0 = 4 \Box_\eta\Xi=0$ and using (\ref{44b}) we get $(1+\eta\cdot\partial_\eta)\Xi=0$ which implies that $\Xi=0$. Taking now a $\epsilon_0$ transformation in (\ref{45}) yields  $\eta\cdot\partial_x\Box_\eta \epsilon_0 =0$ meaning that $\Box_\eta \epsilon_0 = 0$, that is, $\epsilon_0$ is traceless. We have then completely fixed the $\Xi$ transformation and obtained the double traceless condition on the fields and the traceless condition on the gauge parameter needed to describe a HS theory. 

We now expand $\psi_0$ as in (\ref{5a}) to get from (\ref{46}) that
\begin{equation}\label{53}
	\sum_{n=0}^\infty \frac{1}{n!} \, {\eta}^{\mu_1} \dots {\eta}^{\mu_n} \left[ \Box_x \psi^{(0,n)}_{\mu_1\dots\mu_n} - n \partial_{\mu_1} \partial\cdot\psi^{(0,n)}_{\mu_2\dots\mu_n} + \frac{1}{2} n(n-1) \partial_{\mu_1}\partial_{\mu_2} \psi^{(0,n)^\prime}_{\mu_3\dots\mu_n}\right] = 0,
\end{equation}
where a prime denotes contraction of two indices. These are the Fronsdal equations for all integer higher spins. The gauge transformation for $\psi_0$ in (\ref{44}) is the usual gauge transformation for HS fields. We have then obtained the double traceless condition of $\psi^{(0,n)}(x)$ as a field equation while the traceless condition on the gauge parameter appears as a consistency condition for the choice (\ref{45}). These equations were also obtained in \cite{Bekaert:2005in} where they were derived from Fronsdal equations but no action was provided. It should be remarked that \cite{Bekaert:2005in} considered only one helicity in $\psi_0$ but as we have shown here it can be extended to any number of fields in $\psi_0$. To single out just one helicity $s$ we have to impose one further condition 
\begin{equation}\label{57}
	(\eta\cdot\partial_\eta - s)\psi_0=0.
\end{equation}
This selects the term $\psi^{(0,s)}_{\mu_1\cdots\mu_s}$ in $\psi_0$. Gauge invariance now requires that $(\eta\cdot\partial_\eta - s + 1)\epsilon_0 = 0$ so that $\epsilon_0$ has just one component of rank $s-1$ as expected. 

Alternatively, we could go back to the original action (\ref{1}) or to (\ref{1a}) and set $\rho=0$ to get 
\begin{equation}\label{54}
	S = \frac{1}{2} \int d^4 x \,\, d^4\eta  \left[ \delta^\prime(\eta^2+1) (\partial_x \psi_0)^2 + \nonumber 
	 \frac{1}{2} \delta(\eta^2+1) \left( (\partial_\eta\cdot\partial_x \psi_0 + 2 \eta\cdot\partial_x \psi_1)^2 - 4 \partial_x \psi_0 \cdot \partial_x \psi_1 \right) \right].
\end{equation}
It is invariant under (\ref{2.11}) and (\ref{2.12}) with $\rho=0$. The equations of motion from the variation of $\psi_0$ gives (\ref{42}) while the variation of $\psi_1$ gives (\ref{42a}). There are no constraints on the fields. This action is identical to the action (52) of \cite{Segal:2001qq}, which describes a HS theory for all helicities, if we consider the case of flat spacetime and if we identify their fields $h_1$ and $h_2$ as $h_1=\psi_0/\sqrt{2}$ and $h_2=-2 \psi_1/\sqrt{2}$, respectively, when $\mu^2=1$. The symmetries (11) of \cite{Segal:2001qq} are precisely (\ref{2.11}) and (\ref{2.12}) with $\chi_0=0$ so the $\theta$ symmetry is fixed and the transformations are no longer reducible. In a sense, (\ref{1a}) is the generalization of \cite{Segal:2001qq} to the continuous spin case. 

We now implement the choice (\ref{45}) in the action to get 
\begin{eqnarray}\label{55}
	S = &&\frac{1}{2} \int d^4 x \,\, d^4\eta  \left[ \delta^\prime(\eta^2+1) (\partial_x \psi_0)^2 + \nonumber \right. \\
	&&\left. \frac{1}{2} \delta(\eta^2+1) \left( (\partial_\eta\cdot\partial_x \psi_0 - \frac{1}{2} \eta\cdot\partial_x \Box_\eta \psi_0)^2 + \partial_x \psi_0 \cdot \partial_x \Box_\eta \psi_0 \right) \right].
\end{eqnarray}
Now the action is no longer gauge invariant. The variation of the action is proportional to a term depending on $\psi_0$ (and its derivatives) multiplied by $\eta\cdot\partial_x \Box_\eta \epsilon_0$ so that we regain gauge invariance if the gauge parameter is traceless. 
Because of the presence of $\Box_\eta^2$ in the action the equations of motion will have terms up to the second derivative of the delta function. This term will give rise to (\ref{46})  while the term with one derivative of the delta function will identically vanish after using (\ref{46}). The term with the delta function without derivatives reduces to $(\eta\cdot\partial_x)^2 \Box_\eta^2 \psi_0$  after the use of (\ref{46}) giving rise to the double trace condition on $\psi_0$. At the end we get again the Fronsdal action for all spins. 

A third way to proceed is to use the expansion (\ref{5a}) for $\psi_0$ directly in the action (\ref{55}). The $\eta$ integration is divergent since it has to be performed on the hyperboloid enforced by the delta functions. We can perform a Wick rotation $\eta^0 \rightarrow i \eta^0$ so that the integration is now done on the sphere. Alternatively we could have started with the Euclidean version of (\ref{1}) so that the delta functions enforce an integration over a sphere\footnote{Other regularizations are also discussed in \cite{Schuster:2014hca}. Since we get Fronsdal field equations for all spins from (\ref{55}) the regularization of the action must not be a fundamental problem.}. Anyway, since the action is quadratic in $\psi^{(0,n)}(x)$ there will be contributions involving fields of different ranks. If the sum (or difference) of the ranks is odd there will appear an odd number of $\eta$'s so that the integral vanishes. When the sum (or difference) of the ranks is even there appears two equal terms with opposite signs so that they cancel out. Then the action reduces to a sum of quadratic terms with fields of the same rank. The integration on $\eta$ can be performed and we get
\begin{equation}\label{56}
	S = - \pi^2 \sum_{s=0}^\infty \frac{1}{2^s (s!)^2} S_s^{(F)},
\end{equation}
where $S_s^{(F)}$ is the Fronsdal action for helicity $s$. 

Again, in order to get a description for just one helicity $s$ in (\ref{56}) we can impose (\ref{57}) to pick up the helicity $s$ component. We could try to implement this condition   directly into the action through a Lagrange multiplier but this seems very difficult to be accomplished since this condition is not gauge invariant.

\section{Oscillator Basis}\label{s4}

Another approach to handle the tensor indices in HS field theory is through the use of oscillators instead of extra coordinates. The version developed in \cite{Metsaev:2011iz} for flat space makes use of creation and annihilation operators, $\alpha^\mu, \overline{\alpha}^\mu$ satisfying the usual commutation relations $[\overline{\alpha}^\mu,\alpha^\nu]=\eta^{\mu\nu}$ with $\overline{\alpha}^\mu = (\alpha^\mu)^\dagger$. A totally symmetric field of helicity $s$, $\phi_{\mu_1\cdots\mu_s}(x)$, is saturated with the creation operators to form the ket
\begin{equation}\label{58}
	|\phi> = \frac{1}{s!} \alpha^{\mu_1} \dots \alpha^{\mu_s} \phi_{\mu_1 \dots \mu_s}(x) |0>.  
\end{equation}
An action, which reduces to the Fronsdal action, is then written as
\begin{eqnarray}\label{59}
	S^{(M)} = &&- \frac{1}{2} \int d^4x  \left[   < \partial^\mu \phi | ( 1 - \frac{1}{4} \alpha^2 \overline{\alpha}^2 ) | \partial_\mu \phi> - \right. \nonumber \\
 && \left.   < (\overline{\alpha} \cdot \partial_x - \frac{1}{2} \alpha \cdot \partial_x \overline{\alpha}^2) \phi | (\overline{\alpha} \cdot \partial_x - \frac{1}{2} \alpha \cdot \partial_x \overline{\alpha}^2) \phi > \right],
\end{eqnarray}
where $<\phi| = (|\phi>)^\dagger$. The double traceless condition is now $(\overline{\alpha}^2)^2 |\phi> =0$, while the gauge symmetry has the form
\begin{eqnarray}\label{60}
\delta |\phi> &=& \alpha \cdot \partial_x |\epsilon>, \\
|\epsilon> &=& \frac{1}{(s-1)!} \alpha^{\mu_1} \dots \alpha^{\mu_{s-1}} \epsilon_{\mu_1 \dots \mu_{s-1}}(x) |0>,
\end{eqnarray}
with the traceless condition $\overline{\alpha}^2 |\epsilon>=0$. We can find explicitly the relation among the terms in the action (\ref{59}) and the action (\ref{55}) with $\psi_0$ satisfying (\ref{57}) to have just one  helicity. The result is 
\begin{eqnarray}\label{61}
	 <\partial^\mu \phi | \partial_\mu \phi> &&= \frac{2^s s!}{\pi^2} \int d^4 \eta \left[ \delta^\prime(\eta^2+1) (\partial_x \psi_0)^2 + \frac{1}{4} \delta(\eta^2+1) \partial_x \psi_0 \cdot \Box_\eta \partial_x \psi_0 \right], \\
	< \overline{\alpha}^2 \partial^\mu \phi|  \overline{\alpha}^2 \partial_\mu \phi> && = -\frac{2^s s!}{\pi^2} \int d^4 \eta \delta(\eta^2 + 1) \partial_x \psi_0 \cdot \Box_\eta \partial_x \psi_0, \\
   < \overline{\alpha}\cdot\partial_x \phi| \overline{\alpha}\cdot\partial_x \phi> 	&&= \frac{2^s s!}{\pi^2} \int d^4 \eta  \delta(\eta^2 + 1) \left[ -\frac{1}{2} (\partial_\eta \cdot \partial_x \psi_0)^2 + \frac{1}{16} (\eta \cdot \partial_x \Box_\eta \psi_0)^2  \right.\nonumber \\
	&&   \left. -\frac{1}{8} \partial_x \psi_0 \cdot \Box_\eta \partial_x \psi_0 \right], \\
    <\overline{\alpha}\cdot\partial_x \phi| \alpha\cdot\partial_x \,\, \overline{\alpha}^2 \phi> && = \frac{2^s s!}{\pi^2} \int d^4 \eta  \delta(\eta^2 + 1) \left[ -\frac{1}{2} \partial_\eta\cdot\partial_x \psi_0 \,\,\, \eta\cdot\partial_x  \Box_\eta \psi_0    \right.\nonumber \\ 
	&&   \left. + \frac{1}{8} (\eta\cdot\partial_x \Box_\eta \psi_0)^2 - \frac{1}{4} \partial_x\psi_0 \cdot \Box_\eta \partial_x\psi_0 \right], \\
 <\alpha\cdot\partial_x \,\, \overline{\alpha}^2 \phi| \alpha\cdot\partial_x \,\, \overline{\alpha}^2 \phi>  &&=  \frac{2^s s!}{\pi^2}\int d^4 \eta  \delta(\eta^2 + 1) \left[ -\frac{1}{2} \partial_x \psi_0 \cdot \Box_\eta \partial_x \psi_0 - \frac{1}{4} (\eta\cdot\partial_x \Box_\eta \psi_0)^2 \right].
\end{eqnarray}
However it is not apparent how the two set of variables are connected. If there exists a connection between the extra coordinates $\eta$ and the oscillators $\alpha,\overline{\alpha}$ it is not a simple one. Also it is not clear how to lift the trace constraints using the oscillator formalism so that a connection with our formulation with two fields is not apparent. 

A similar situation happens with the Francia-Sagnotti formulation in \cite{Francia:2002aa}. The trace constraint on the completely symmetric tensor for a given helicity is lifted with two completely symmetric gauge fields of rank $s-3$ and $s-4$. Their action contains higher spacetime derivatives while ours has only two spacetime derivatives so it seems that there is no clear connection between the two formalisms as well.

A more direct connection between the original Schuster-Toro action (\ref{1}) with $\rho=0$ and the oscillator formalism  was provided in \cite{Schuster:2014hca}. The integral over $\eta$-space was performed using a star product which is equivalent to take $\eta^\mu$ and $\partial/\partial{\eta^\mu}$ as creation and annihilation operators. Then a correspondence can be found between each term of (\ref{1}) with each term of the Fronsdal action. It is also easily found that the $\delta^\prime$ term of (\ref{1}) corresponds to the first term of (\ref{59}) while the $\delta$ term corresponds to the second one. 

\section{Physical Contents}\label{pc}

Let us now return to the continuous spin case and analyse the equations of motion (\ref{30}) and (\ref{31}), taking into account both symmetries present in (\ref{2.25}) and (\ref{2.26}), to unveil their physical degrees of freedom and to understand how they do fit into representations of the Poincar\'e group. So let us briefly recollect the main facts about massless irreducible representations of the Poincar\'e group. To this end we have to find out how the Pauli-Lubansky operator $W^\mu = \epsilon^{\mu\nu\rho\sigma}P_\nu J_{\rho\sigma}$ acts on the fields. In a light-cone frame where $ds^2 = dx^+ dx^- - (dx^i)^2, (i=1,2)$ and with a light-like momentum with components $k_+\not=0$, $k_- = k^i=0$, the Pauli-Lubansky operator is given by $W^\mu = - i k_+ \epsilon^{+\mu\nu\rho} J_{\nu\rho}$.  We then find that $W^+ = 0, W^- = - i k_+ \epsilon^{ij} J_{ij}$ and $W^i = - i k_+ \epsilon^{ij} J_{j-}$, with $\epsilon^{12}=1$. For massless particles the components of $W^\mu$ satisfy the two dimensional Euclidean space algebra of $E_2$ given by the non vanishing commutators $[h, {\cal W}_\pm] = \pm {\cal W}_\pm$, where  ${\cal W}_\pm = W^1 \pm i W^2$  and $h = i W^-/k_+$ is the helicity operator. The Casimir operator is then $W^2 = - {\cal W}_+ {\cal W}_-$. We can then consider basis vectors which are simultaneously eigenvectors of $W^2$ and $h$, with eigenvalues ${\rho}^2$ and $h$, respectively, 
\begin{eqnarray}
	W^2|{\rho},h> &=& {\rho}^2 |{\rho},h>,  \qquad \quad \,\,\, {\rho}^2 \ge 0, \label{6.a} \\
	h |{\rho},h> &=& h |{\rho},h>, \qquad \qquad h = 0, \pm 1, \pm 2 \dots \label{6.b}
\end{eqnarray}
For ${\rho}^2 = 0$ we have ${\cal W}_\pm |0,h>=0$  and the irreducible representations are one dimensional giving rise to the usual helicity states. For ${\rho}^2 > 0$ we have 
\begin{equation}
	{\cal W}_\pm |{\rho},h> = \mp  {\rho} |{\rho},h \pm 1>, \label{6.c}
\end{equation}
so that ${\cal W}_\pm$ increases/decreases the helicity by one unit leading to a sequence of basis vectors $\{ |{\rho},h>, h = 0, \pm 1, \pm 2, \dots \}$ so the irreducible representations are infinite dimensional. This is a continuous spins representation with continuous spin ${\rho}$. These representations may also be multivalued. Notice that when we take the limit ${\rho} \rightarrow 0$ in the continuous spin case it does not reduce to the ${\rho}=0$ case since we get an infinite number of helicity states in which each helicity appears once. 

Let us first consider the Schuster and Toro formulation. The action of the Pauli-Lubanski vector on $\Psi(\eta,x)$ is given by $W^\mu \Psi = - \epsilon^{\mu\nu\rho\sigma} \partial_{x\mu} \eta_{\rho}\partial_{\eta\sigma} \Psi$ so that 
\begin{equation}
	W^2 \Psi = \left[-\eta\cdot\partial_\eta(1+\eta\cdot\partial_\eta) \Box_x  + \eta^2 \Box_\eta \Box_x  + 2 \eta\cdot\partial_\eta \, \eta\cdot\partial_x \, \partial_\eta\cdot\partial_x  
	 - (\eta\cdot \partial_x)^2 \Box_\eta  - \eta^2 (\partial_x\cdot \partial_\eta)^2\right] \Psi. \label{d6.4}
\end{equation}
Using the field equations we find that
\begin{equation}
	\delta(\eta^2+1) W^2 \Psi = \delta(\eta^2+1) ( \rho^2 \Psi + \delta_\epsilon \Psi), \label{d6.5}
\end{equation}
where $\delta_\epsilon$ means a gauge transformation (\ref{2}) with parameter 
\begin{equation}
\epsilon = (1+\eta\cdot\partial_\eta) (\partial_\eta\cdot\partial_x - \rho)\Psi - (1+ 2\eta\cdot\partial_\eta + (\eta\cdot\partial_\eta)^2 ) \Delta \Psi - (\eta\cdot\partial_x + \Delta)\Box_\eta \Psi. \label{d6.6}
\end{equation}
As expected we get (\ref{6.a}) up to a gauge transformation confirming the results of \cite{Schuster:2014hca} which were obtained after gauge fixing. Notice that we get the same result by computing $W^2 \delta(\eta^2+1)\Psi$ since the delta function is a scalar. We now turn to the formulation (\ref{30},\ref{31}) outside the hyperboloid $\eta^2+1=0$. Now we get
\begin{eqnarray}
	W^2 \psi_0 &=& \rho^2 \psi_0 + \delta_\epsilon\psi_0 + \delta_\Xi \psi_0, \label{d6.7}\\
	W^2 \psi_1 &=& - \eta^2 \rho^2 \psi_1 + \delta_\epsilon\psi_1 + \delta_\Xi \psi_1, \label{d6.8}
\end{eqnarray}
with gauge parameters 
\begin{eqnarray}
	\epsilon_0 &=& - \eta\cdot\partial_\eta(1+\eta\cdot\partial_\eta) \partial_\eta\cdot\partial_x \psi_0 -\rho(2+3\eta\cdot\partial_\eta + (\eta\cdot\partial_\eta)^2 ) \psi_0 + (\eta^2\Delta - \eta\cdot\partial_x)\Box_\eta\psi_0 \nonumber\\
	&&- 2(2+3\eta\cdot\partial_\eta+ (\eta\cdot\partial_\eta)^2 ) \eta\cdot\partial_\eta\psi_1 +2\eta^2(\eta\cdot\partial_x\Box_\eta + 3 \partial_\eta\cdot\partial_x - \rho)\psi_1, \label{d6.9}\\
	\Xi &=& -\rho^2 \psi_0, \label{d6.10}
\end{eqnarray}
for the gauge transformations (\ref{2.25},\ref{2.26}). We then have to go to the hyperboloid $\eta^2+1=0$ in order to have a CSP with continuous spin $\rho$ because of the first term on the RHS of (\ref{d6.8}). Hence CSP's live only on the hyperboloid and not on all of $\eta$-space. 

Having shown that $W^2$ has the expected eigenvalue we now have to make sure that we have the correct physical degrees of freedom. Firstly we will take into account the $\Xi$ symmetry of (\ref{30},\ref{31}). Notice that (\ref{2.26}) allow us to use it to set $\psi_1=0$ so that the $\Xi$ symmetry is fixed. The equations of motion (\ref{30}) and (\ref{31}) then reduce to
\begin{eqnarray}
	(\Box_x - \eta\cdot\partial_x\Delta) \psi_0&=&0, \label{7.2} \\
	\Delta^2 \psi_0 &=& 0. \label{7.3a}
\end{eqnarray}
Applying $\Delta$ to (\ref{7.2}) yields $\eta\cdot\partial_x \Delta^2\psi_0=0$ so that $\Delta^2\psi_0=0$ and (\ref{7.3a}) are not independent. Another consequence of $\psi_1=0$ is that $\Delta\epsilon_0=0$. We now have to deal with the gauge symmetry. A harmonic like gauge is $\Delta \psi_0=0$ which implies that $\Box_x\psi_0=0$ and $\Box_x \epsilon_0=0$. We can then go to momentum space recalling that the only non vanishing component of the momentum is $k_+$. Then the gauge choice $\Delta\psi_0=0$ can be solved for the components of $\tilde{\psi}_0(\eta,k)$ as 
\begin{equation}
	\tilde{\psi}_{0 \underbrace{- \dots -}_{\text{p times}} A_1 \dots A_n}(k) = \left( -\frac{\rho}{2ik_+} \right)^p \tilde{\psi}_{0 A_1 \dots A_n}(k), \quad p\ge 1, \label{7.6}
\end{equation}
where the index $A$ stands for $(+,i)$. This means that the independent components of $\tilde{\psi}_0(\eta,k)$ are $\tilde{\psi}_{0 A_1 \dots A_n}(k)$. Since $\epsilon_0$ satisfies the same equation we get the same result for $\tilde{\epsilon}_0$ so its independent components are $\tilde{\epsilon}_{0 A_1 \dots A_n}(k)$. Having identified the independent components of $\tilde{\psi}_0$ and $\tilde{\epsilon}_0$ we can now use the gauge transformation $\delta\psi_0 = \eta\cdot\partial_x \epsilon_0$ to find out the physical components of $\tilde{\psi}_0$. In components the gauge transformation reads
\begin{equation}
	\delta\tilde{\psi}_{0\mu_1\dots\mu_n}(k) = \frac{1}{(n-1)!} i k_{(\mu_1} \tilde{\epsilon}_{0 \mu_2\dots\mu_n)}(k), \label{7.7}
\end{equation}
resulting in the usual gauge transformation for a symmetric tensor field. Since the only non vanishing component of $k^\mu$ is $k_+$ we can use all components of $\tilde{\epsilon}_0$ to gauge away all the components of $\tilde{\psi}_0$ having one or more + components, so that the gauge invariant components have no +  indices, that is, they are $\tilde{\psi}_{0i_1\dots i_n}$. We get this same result if we take $\rho=0$. In any case $\tilde{\psi}_{0i_1\dots i_n}$ form helicity representations of the Poincar\'{e} group for all integer values of the helicity.  Notice, however, that there is no traceless condition on $\tilde{\psi}_{0i_1\dots i_n}$ so that each helicity is infinitely degenerated. 

Up to now we are outside the $\eta$-hyperboloid. The expansion (\ref{5a}) for $\tilde{\psi}_0$ on the hyperboloid is
\begin{equation}
	\tilde{\psi}_0(\hat{\eta},k) = \sum_{s=0}^{\infty} \frac{1}{s!} \hat{\eta}^{\mu_1} \dots \hat{\eta}^{\mu_s} \tilde{\psi}_{\mu_1\dots\mu_s}(k), \label{d6.15}
\end{equation}
where $\hat{\eta}^\mu=\eta^\mu/|\eta|$ satisfies $\hat{\eta}^2=-1$. Because of this constraint all traces of $\tilde{\psi}_{\mu_1\dots\mu_s}$ can be grouped together so that the irreducible pieces of the expansion (\ref{d6.15}) are traceless, that is,
\begin{equation}
	\tilde{\psi}_0(\hat{\eta},k) = \sum_{s=0}^{\infty} \frac{1}{s!} \hat{\eta}^{\mu_1} \dots \hat{\eta}^{\mu_s} \tilde{\psi}^T_{\mu_1\dots\mu_s}(k), \label{d6.16}
\end{equation}
where $\tilde{\psi}^T_{\mu_1\dots\mu_s}$ is completely traceless. Then on the hyperboloid $\eta^2+1=0$ $\tilde{\psi}_0(\hat{\eta},k)$ has only traceless fields so that it describes all integer helicities each one appearing just once as expected for a CSP. This is also in agreement with the results found in Section \ref{s3} and \cite{Schuster:2014hca} for the HS case since the limit $\rho\rightarrow 0$ is well defined.  

Finally we would like to comment on the gauge choice (\ref{31.a}) made in Section \ref{s6} to get the results of \cite{Bekaert:2005in}. Now $\psi_0$ vanishes on the $\eta$ hyperboloid so we will keep $\psi_1$ as our independent variable. In this gauge the equations of motion (\ref{30}) and (\ref{31}) become
\begin{eqnarray}
	&(\eta^2+1) ( \Box_x - \eta\cdot\partial_x \Delta) \psi_1 - 4 (\eta\cdot\partial_x)^2 \psi_1 = 0, \label{7.17} \\
	&( \Box_x + \eta\cdot\partial_x \Delta) \psi_1 + \frac{1}{6} (\eta^2+1) \Delta^2 \psi_1=0, \label{7.18}
\end{eqnarray}
while the local symmetry implies that 
\begin{equation}
	\eta\cdot\partial_x \epsilon_0 + \frac{1}{2} (\eta^2+1) \Delta \epsilon_0=0. \label{7.18a}
\end{equation}
The harmonic gauge is now 
\begin{equation}
	(\eta^2+1) \Delta\psi_1 + 4\eta\cdot\partial_x \psi_1=0. \label{7.19}
\end{equation}
The equations of motion (\ref{7.17}) and (\ref{7.18}) gives $\Box_x \psi_1=0$ while the gauge transformation of $\psi_1$ applied in (\ref{7.19}) together with (\ref{7.18a}) imply in $\Box_x\epsilon_0=0$. We can now solve (\ref{7.18a}) to find out the independent components of $\epsilon_0$ in momentum space and they are $\tilde{\epsilon}_{0+\dots +i_1\dots i_n}$ with any number of $+$ indices. We can also solve (\ref{7.19}) to find out the independent components of $\psi_1$ and they are $\tilde{\psi}_{1+\dots +i_1\dots i_n}$ with any number of $+$ indices. Then, using the gauge transformation in (\ref{2.26}) we find that all $\tilde{\psi}_{1+\dots +i_1\dots i_n}$ with one or more $+$ indices can be gauged away so that all the $\tilde{\epsilon}_{0+\dots +i_1\dots i_n}$ are used. The remaining components $\tilde{\psi}_{1 i_1\dots i_n}$ are gauge invariant and no traceless condition is found. Going to the hyperboloid $\eta^2+1=0$ we find that $\tilde{\psi}_{1 i_1\dots i_n}$ becomes traceless and we get the same physical degrees of freedom as in the $\psi_1=0$ gauge. In this way we have confirmed that different gauge choices lead to the same degrees of freedom and to the same CSP.

\section{Conclusions}\label{conclusions}

We have shown that there is a rich structure behind the Schuster-Toro action. The Schuster-Toro local transformations are reducible and after the expansion of $\Psi(\eta,x)$ in powers of $\eta^2+1$ reducibility was used to eliminate all components of the gauge parameter $\epsilon$ except the first one while the $\chi$ symmetry was responsible for the elimination of all terms of $\Psi$ except the first two. The reduced action depends on two fields $\psi_0$ and $\psi_1$ and is invariant under reducible but simpler local transformations. As shown here it reproduces the equations found by \cite{Bekaert:2005in} for  the continuous spin case and the action of \cite{Segal:2001qq} in the HS case giving rise to an alternative formulation for CSPs. 

The extension of our results to dimensions other than four is straightforward. The most important extension right now is the addition of possible deformations of the gauge symmetry to include self-interactions. The coupling to gravity in particular deserves attention since it may provide an alternative way to look for CSPs and HS interactions in non-flat backgrounds, in particular in AdS. 
%It will be also very interesting to investigate if the equivalence of theories living in different hyperboloids will persist when interactions are turned on. 

It is known that in the context of $AdS_4/CFT_3$ some critical models in the boundary are dual to higher spin fields in $AdS_4$ \cite{Klebanov:2002ja}. Such higher spin theories have been extensively studied (for a recent review see \cite{Didenko:2014dwa}) and only recently an action has been proposed \cite{Boulanger:2011dd,Boulanger:2012bj}. The new action  presented here may be an alternative to the study of this duality. Another situation where the HS results found here can be applied is in the small tension limit of string theory \cite{Savvidy:2003fx,Mourad:2006xk}.

\acknowledgments

I would like to thank Andrei Mikhailov for conversations and Xavier Bekaert for pointing out reference \cite{Segal:2001qq}. This work is supported by CNPq grant 304116/2010-6 and FAPESP grant 2012/51444-3.

\end{document}